\documentclass[twocolumn]{svjour3}          
\pdfoutput=1
\usepackage{mathptmx}      

\usepackage{fullpage}
\usepackage{amssymb}
\usepackage{breakcites}
\usepackage[pdftex]{graphicx}
\usepackage{float}
\usepackage[toc,page]{appendix}
\usepackage{mathtools}

\usepackage{algorithm}
\usepackage{algpseudocode}

\usepackage{microtype}
\usepackage{rotating}
\begin{document}

\title{Fast implementation of a Bayesian unsupervised segmentation algorithm
\thanks{This work was partly supported by FAPESP grants 16/02175-0 and CNPq grant 308405/2013-7. The authors are grateful for the support received from IME-USP, the Institute of Mathematics and Statistics of the University of Sao Paulo;  FAPESP - the State of S\~{a}o Paulo Research Foundation (grants CEPID-CeMEAI 2013/07375-0 and CEPID-Shell-RCGI 2014/50279-4); and CNPq - the Brazilian National Counsel of Technological and Scientific Development (grant PQ 301206/2011-2).The authors acknowledge the Laje de Santos Marine State Park Team for supporting this project.}
}

\titlerunning{Fast implementation of a Bayesian unsupervised segmentation algorithm}     

\author{Paulo Hubert $^1$ \and Linilson Padovese $^2$ \and Julio M. Stern $^3$ }

\authorrunning{Hubert, P., Padovese, L., Stern, J.M.} 

\institute{$^1$ \at
              Rua do Mat\~{a}o, 1010 - Butant\~{a}, S\~{a}o Paulo - SP, 05508-010, Brazil \\
              Tel.: +55-11-96040-3161\\
              \email{phubert@ime.usp.br}           
           \and
           $^2$ \at
              Av. Prof. Luciano Gualberto, 380 - Butant\~{a}, S\~{a}o Paulo - SP, 05508-010, Brazil \\
            \and
            $^3$ \at
              Rua do Mat\~{a}o, 1010 - Butant\~{a}, S\~{a}o Paulo - SP, 05508-010, Brazil \\
              Tel.: +55-11-96040-3161
}

\maketitle

\begin{abstract}
In a recent paper, we have proposed an unsupervised algorithm for audio signal segmentation entirely based on Bayesian methods. In its first implementation, however, the method showed poor computational performance. In this paper we address this question by describing a fast parallel implementation using the Cython library for Python; we use open GSL methods for standard mathematical functions, and the OpenMP framework for parallelization. We also offer a detailed analysis on the sensibility of the algorithm to its different parameters, and show its application to real-life subacquatic signals obtained off the brazilian South coast. Our code and data are available freely on github.
\keywords{Signal segmentation \and Unsupervised learning \and Bayesian inference \and Numerical computation}

\end{abstract}

\section{Introduction}
\label{intro}

The problem of signal segmentation arises in different contexts (\cite{Makowsky2014}  \cite{Ukil2006} \cite{Schwartzman2011} \cite{Kuntamalla2014} \cite{Theodorou2014}). The problem is broadly defined as follows: given a discretely sampled signal $y \in \Re^N$, divide it in contiguous sections that are internally homogeneous with respect to some characteristic. The segmentation is thus based on the premise that the signal structure changes one or many times during the entire sampled period $T = N / f_s$ (where $f_s$ is the sampling rate in $Hz$), and one is looking for the \emph{change points}.

The signal structure, in this context, refers to a parametric model describing the signal evolution; this can be a tonal model, for instance, where one is interested in detecting changes on the fundamental frequency or other features of the spectrum. 

In a previous paper \cite{Hubert2018} we introduced a probabilistic model where the changing points refer to changes in the signal's average power; our main goal was to segmentate acoustic subacquatic signals, in order to separate sections that are likely to contain significant events (boat engines, fish choruses, whales, etc.). We compared our algorithm to a traditional peak detection strategy, and found promising results.

Our algorithm, however, is very computationally intensive, and in our first tests, using a pilot version written in MATLAB \textregistered, the algorithm took on average $120$ seconds to completely segmentate a $15$ minutes audio signal with many power changes. To improve this situation, we present here a Python version of the algorithm, using the Cython package \cite{Cython} to generate optimized C code.

With this new, optimized version of the algorithm, we offer a detailed analysis of the algorithm's sensibility to the different parameters and in different situations. Finally, we illustrate the algorithm's performance in samples of subacquatic recordings in the region of Santos, SP, Brazil. Both the algorithm and the samples are openly available at \url{https://github.com/paulohubert/bayeseg}.

The paper is organized as follows: the following section reviews the motivation, the model and the sequential segmentation algorithm. Section \ref{sec:impl} discusses in detail the fast implementation using Cython. In section \ref{sec:param} we study the algorithm's sensibility to changes in the parameters, in different situations and using simulated signals. Section \ref{sec:app} presents an analysis of real subacquatic signals. Section \ref{sec:conc} concludes the paper.

\section{The sequential segmentation algorithm}\label{sec:algo}

\subsection{Motivation}\label{subsec:motiv}

Our main goal when developing this algorithm was to analyze signals coming from the OceanPod: an acoustic underwater recording system developed by the engineering school at University of São Paulo (\emph{EP - USP}) \cite{OceanPod1}. 

One OceanPod was installed in January 2015 at the \emph{Parque Estadual Marinho da Laje de Santos} (Laje de Santos Marine State Park), a marine preservation park in the region of Santos, SP, Brazil. The prime motivation for installing the hydrophone in this location was to study the patterns of behavior of fishing vessels that navigate through the park. Since fishing in the park is prohibited, understanding of these patterns can be a valuable asset in the design of fiscalization policies. 

As a secondary objective, the data coming from the hydrophone also allows the study of other noise irradiating events, either caused by man or nature. Obtaining samples of different kinds of acoustic events from a region with high diversity of marine life can be potentially very useful for the investigation of animal behavior off the brazilian coast, in the least by providing annotated samples to be fed to classification algorithms that can later be developed to search for specific events.

The OceanPod is installed at a depth of $20$ m, and delivers $15$ minutes long audio files sampled at $11,205 Hz$. With this data, we are interested in indicating times and duration of boats passing by, but also in detecting other possibly interesting events. Given this broad objective of separating segments regardless of their specific nature, we developed a methodology to divide the signal into segments based on differences of the average power only. This leads us to the sequential segmentation algorithm, which we present next.

\subsection{Model and algorithm}\label{subsec:model}

To define the sequential segmentation algorithm (\emph{SeqSeg}), we start by assuming that the (discretely sampled) signal $y \in \Re^N$ has $0$ mean amplitude, and finite power $\sigma_0^2$. Given these two assumptions, by the maximum entropy principle \cite{Jaynes1982} \cite{Jaynes1987} we adopt a Gaussian probabilistic model for the signal, $y_t \sim \mathcal{N}(0, \sigma_0^2)$.

Now suppose that from sample $t = \bar{t}$ on, the signal's power shifts from $\sigma_0^2$ to $\sigma_1^2 = \delta \cdot \sigma_0^2$, with $\delta > 0$. This leads to the following probabilistic model:

\begin{equation}\label{model}
y_t \sim
  \begin{cases}
   \mathcal{N}(0, \sigma_0^2)&\quad\text{if } t \le \bar{t}\\
   \mathcal{N}(0, \delta\sigma_0^2)&\quad\text{if } t > \bar{t}\\
  \end{cases}
\end{equation}

The likelihood function associated with this model is thus

\begin{align}\label{verot}
 \mathcal{L}(\bar{t}, \sigma_0^2, \delta | y)& = (2\pi\sigma_0^2)^{-\frac{N}{2}}\delta^{-\frac{N-\bar{t}}{2}} \times \\
 & exp\left[-\frac{\sum_{t=1}^{\bar{t}}y_t^2}{2\sigma_0^2}-\frac{\sum_{t=\bar{t}+1}^Ny_t^2}{2\delta\sigma_0^2}\right]
\end{align}

By adopting a uniform, uninformative prior for $\delta$, and Jeffreys's prior \cite{Jaynes1968} \cite{Jeffreys1946} $\pi(\sigma_0) = 1/\sigma_0$ for $\sigma_0$, we can integrate equation \ref{verot} analitically and obtain the posterior

\begin{align}\label{postt}
 P(\bar{t} | y) \propto & \pi(\bar{t})\cdot\left(\sum_{t=1}^{\bar{t}}y_t^2\right)^{-\frac{(\bar{t}+6)}{2}} \left(\sum_{t=\bar{t}+1}^Ny_t^2\right)^{-\frac{(N-\bar{t}-6)}{2}} \times \\
 & \Gamma\left(\frac{\bar{t}+6}{2}\right)\Gamma\left(\frac{N-\bar{t}-2}{2}\right)
\end{align}

This is a discrete distribution with support on $\{3, N-3\}$; to obtain the MAP (\emph{Maximum a posteriori}) estimate for $\bar{t}$, we can simply evaluate the above function on its entire domain. 

Our real signals, however, will usually have more than one change point. To model this situation, one approach would be to define the model for $k$ change points, given knowledge of $k$. This, however, would insert many complications in the algorithm design. For instance, there would be the problem of estimating $k$ when it is unknown (the most frequent situation), and the support of the posterior distribution for the change points would be many-dimensional, making it much more difficul to find the MAP estimates.

To avoid these problems, we propose to apply the algorithm in a sequential manner: we start by dividing the original signal in two sections, using model \ref{postt} above. Then, \emph{if there is sufficient evidence of a difference in power between the two segments}, we apply again the same method to each section, and so on until no more significantly different segments are found. As we will see on the following sections, this sequential approach is efficient and captures precisely the changepoints even when there are many of them in the signal.

\begin{figure}[H]
 \centering
 \includegraphics[width=0.5\textwidth]{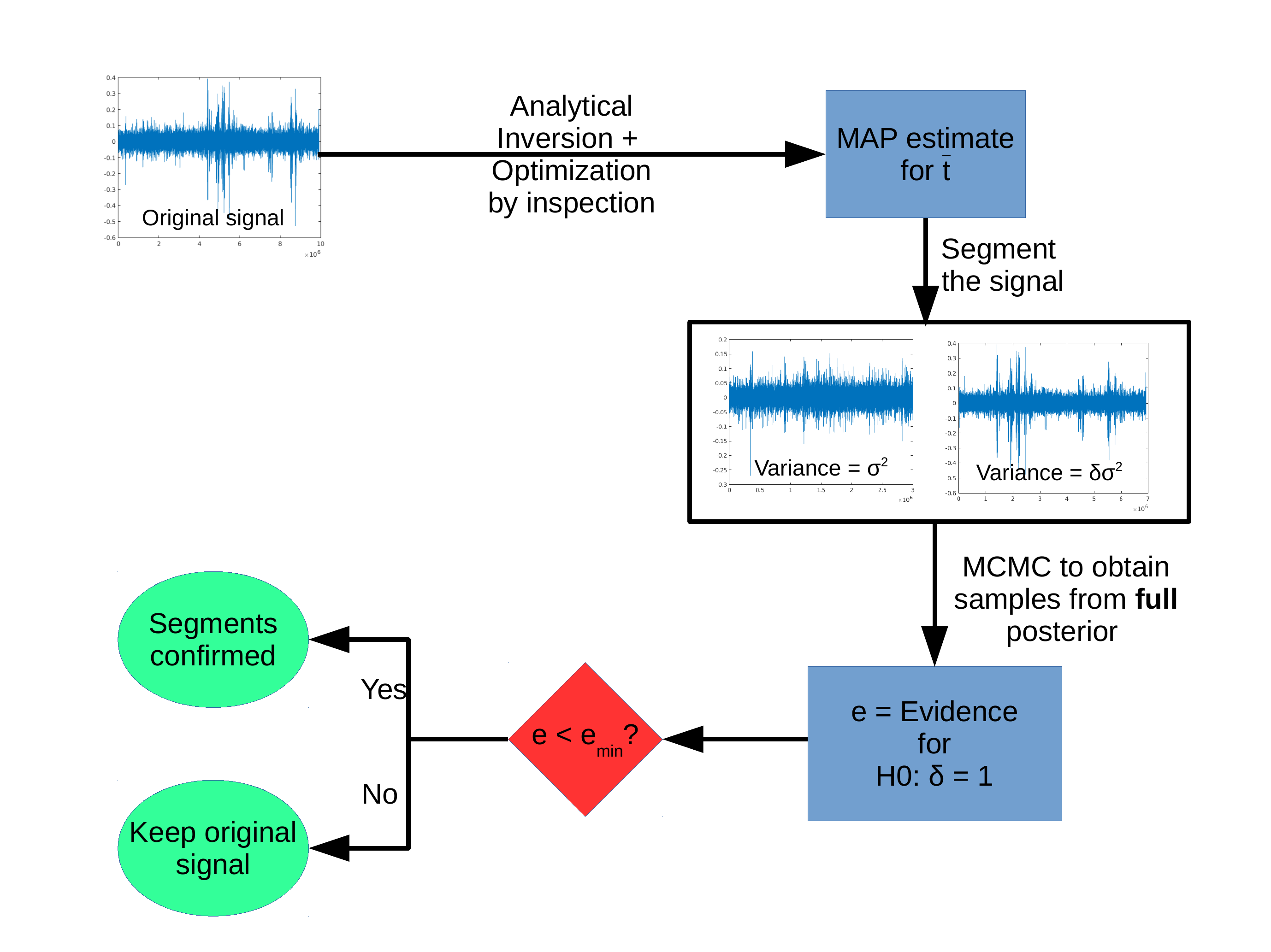}
 \caption{One step of the sequential segmentation algorithm.}
 \label{fig:diag}
\end{figure}

With this approach, one full step of the algorithm will consist of two substeps: first, to estimate the segmentation point $\bar{t}$; second, to compare the power of the segments, calculating a measure of evidence for the hypothesis $H_0:\delta = 1$. A diagram illustrating the algorithm's structure is included in figure \ref{fig:diag}.

The evidence measure we use is the \emph{Full bayesian significance test} (FBST) of Pereira and Stern \cite{Pereira1999}. This is an inferential procedure aimed at the calculation of evidence for \emph{sharp hypothesis} (hypothesis that induce a reduction on the dimension of the parametric space). The hypothesis $H_0:\delta = 0$ is sharp, since under $H_0$ we have the parametric space $\Theta_0 = \Re_+$, and the unrestricted (full) parametric space is $\Theta = \Re_+ \times \Re_+$.

To compare the segments with respect to their variances, we apply again the Gaussian model, now conditionally on $\bar{t}$. This means that the likelihood on equation \ref{verot} is the same, but now $\bar{t}$ is considered fixed (known). Given the prior distributions $\pi_\sigma$ and $\pi_\delta$, and defining the segments $y^1 = \{y_t : t \le \bar{t}\}$ and $y^2 = \{y_t : t > \bar{t}\}$, the full posterior is given by

\begin{align}\label{eq:fullpost}
 P(d, s | y^1, y^2) \propto & \pi_\delta(d)\pi_\sigma(s)(2\pi s^2)^{-\frac{n_1+n_2}{2}} d^{-\frac{n_2}{2}} \times \\
 & exp\left[-\frac{\sum_{t=1}^{n_1}y_{1,t}^2}{2s^2}-\frac{\sum_{t=1}^{n_2}y_{2,t}^2}{2ds^2}\right]
\end{align}

where $n_1$ and $n_2$ are the segments' sizes. 

We adopt again a Jeffreys' prior for $\sigma_0$. For $\delta$, however, in order to be able to control the sensibility of the algorithm, we propose a Laplace prior distribution centered around $d = 1$:

\begin{equation}
 \pi_\delta(d) = \frac{1}{\beta}e^{-\frac{|d-1|}{\beta}}
\end{equation}

The main reason for using the Laplace distribution is that it is a) symmetric, b) high-peaked around its mean, c) has higher tails than the gaussian. Also, by carefully choosing the hyperparameter $\beta$, we hope to be able do adapt our algorithm to different situations. In section \ref{sec:param} below we investigate the algorithm results under different choices for $\beta$.

Calculation of evidence in the FBST framework now involves two steps: first, we obtain the maximum posterior under $H_0$:

\begin{equation}
 p_0 = max_{s}P(1, s | y^1, y^2)
\end{equation}

In our model, this can be done analytically; by differentiating equation \ref{eq:fullpost} we find

\begin{equation}
 s_0 = argmax_s P(1, s | y^1, y^2) = \frac{\sum_{i=1}^N y_i^2}{N + 1}
\end{equation}

After obtaining $p_0$, we now define the \emph{evidence against $H_0$} as

\begin{align}
 T(p_0) = & \{(d, s) \in \Re_+ \times \Re_+ : P(d,s|y^1, y^2) > p_0 \\
 ev(H_0) = & \int_{T(p_0)} P(d, s | y^1, y^2) ds dd
\end{align}

The evidence against the hypothesis is the posterior measure of the \emph{surprise set} of points (in the full parametric space) having larger posterior values than the maximum posterior under $H_0$. If $ev(H_0)$ is high (i.e., the surprise set has high measure), the manifold defined by $H_0$ traverses regions of low posterior probability, leading us to assign large evidence against the hypothesis.

To obtain the evidence against $H_0$, thus, we must estimate the integral above. This cannot be done analytically, and simulation methods must be adopted.

We choose to apply the Adaptive Metropolis algorithm of Haario \emph{et al} \cite{Haario2001}. This is a random walk, Metropolis algorithm with Gaussian candidate distributions for the increment in the parameters. To improve the mixing properties of the chain, the covariance matrix of the candidate distribution is first estimated from a few runs of the usual Metropolis algorithm, and then adaptively updated during the burn-in period of the chain. In the next section, we describe the details of the algorithm implementation.

\section{Fast implementation using Cython}\label{sec:impl}

To implement the SeqSeg algorithm we choose the python language, for a few reasons. First, because python is now widely used in the scientific community (specially after the release of NumPy and SciPy), and is thus a convenient programming language to use for open source scientific software. Second, because it is a simple to read language, that will make it easier for users to understand and modify the code. And third, because of the Cython package.

Cython (\url{http://cython.org}) is a package that originated from the Pyrex project \cite{Cython}. Its objective is to automatically generate optimized C code from a python source. To accomplish this, the package demands only that the programmer explicitly declares and types each variable in the Python code; by running the Cython interpreter, and compiling the resulting C code, one is able to improve numerical performance a hundred (sometimes a thousand) fold. Also, Cython supports the OpenMP (\url{http://www.openmp.org}) library for multiprocessing programming, with allows the programmer to boost even further the numerical performance of parallelizable methods.

Besides using Cython and parallelizing our algorithm, we also chose to adopt the GSL library (\url{http://www.gnu.org/software/gsl}) for random number generation, and the standard C implementations of \textit{math.h} for the mathematical functions (such as the gamma and log-gamma functions). All of these components together allowed us to improve our average computing time from $120s$ to less than $1$ second.

In the remainder of this section, we introduce our \textit{bayeseg} python module, which implements the SeqSeg class. SeqSeg is the interface that allows easy application of our algorithm. We also describe the details involved in the implementation of the brute-force optimization step, and the calculation of the evidence value.

\subsection{The \textit{bayeseg} module}\label{subsec:bayeseg}

We implement our algorithm in a Python module which we call \textit{bayeseg} (from Bayesian segmentation). The code is avaliable on the git repository \url{http:/github.com/xxxx}, along with the real data samples used in this paper. 

The module is composed of two classes: the \emph{OceanPod} class is a simple interface that allows the user to easily read the audio files from the hydrophone. It also can convert segments' indexes to \emph{datetime} objects, and vice-versa, based on the file name (assuming a specific pattern of file name that contains the date and time of the recording). 

The second class is the \emph{SeqSeg} class, that implements the algorithm. In the next section, we describe this class in more detail.

\subsection{The SeqSeg class}\label{subseg:seqseg}

Our interface class, \textit{SeqSeg}, provides methods that allow the user to load data, initialize the parameters, and obtain the segmentation of a signal.

The first method, \textit{feed\_data}, takes a NumPy array as argument, stores it internally, and preprocess the signal calculating and storing all the cumulative sums in the form $\sum_{i=1}^N y_i^2$. This calculation speeds up the segmentation process by eliminating the need to iterate through the vector calculating the sum of squares at each step.

The \textit{initialize} method of SeqSeg takes the following arguments:

\begin{itemize}
 \item \textit{double} $\beta > 0$: the hyperparameter of the prior for $\delta$;
 \item \textit{double} $0 < \alpha < 1$: the threshold for the decision functin; when comparing two segments, the algorithm will accept that they have significantly different powers whenever the evidence for $H_0:\delta = 1$ is higher than $\alpha$. In other words, the higher the value of $\alpha$, the greater the number of segments produced by the algorithm;
 \item \textit{int mciter}: the number of iterations for the MCMC algorithm;
 \item \textit{int mcburn}: the number of iterations for the burn-in period of the MCMC algorithm;
 \item \textit{int nchains}: the number of parallel chains to run.
\end{itemize}

There are two more quantities to fully define the SeqSeg algorithm: the minimum segment length and the time resolution. 

The minimum segment length ensures that the algorithm stops; it imposes a minimum length for a valid segment, and whenever the change point estimation step finds a cutting point that creates segments with a lower length than this minimum, it skips the evidence calculation step, stopping the segmentation.

The time resolution affects the change point estimation step only, in the following way: with a time resolution of $l$, the brute-force MAP estimation step will evaluate the posterior for $\bar{t}$ on the points $\{3 + i\cdot l, i = 0,..., (N-6)/l\}$. A time resolution greater than 1 speeds up the algorithm (since the brute-force MAP estimation of $\bar{t}$ will evaluate the objetive function only on points separated by $l$) at the cost of possibly missing a change point.

Both of these parameters can be set on the run, when calling the \textit{segments} method. This method takes as parameters the minimum segment length and the time resolution, and returns a list with the index of all change points as estimated by the SeqSeq algorithm.

\subsection{Estimation of the segmentation point}\label{subsec:estimation}

The first step of the SeqSeg algorithm is the MAP estimation of the change point. This is accomplished by simply evaluating the posterior \ref{postt} over the set of points $\{3 + i\cdot l, i = 0,..., (N-6)/l\}$, where $l$ is the time resolution.

This step can be very demanding, specially with large signals. It involves the calculation of the log-gamma function twice, calculation of two natural logarithms, plus 21 floating point operations. 

To achieve maximum performance, and also to make this step parallel, we implement the posterior \ref{postt} as a \textit{cdef} function (named \textit{cposterior\_t} in our code). In Cython, this means that this function will not be available from the Python module, but will be restricted to the C compiled library. In our python implementation, this function is also defined as \emph{nogil}, which means that it does not make explicit use of the Python interpreter. This is necessary to allow for parallelization (Python is defined to not allow two or more threads of the interpreter to run simultaneously, a feature known as \emph{global interpreter lock, gil}; to write parallel code, then, one has to circumvent this by explicitly declaring the function to be interpreter-independent), and the consequence is that the code for the function can not use any Python object; it is restricted to basic operations, and to calls to other \emph{nogil} functions.

The parallelization is implemented using the \textit{prange} function from Cython; \textit{prange}, according to the Cython online documentation, causes OpenMP to automatically start a thread pool and distribute the work according to a defined schedule (which can be set to \emph{static}, \emph{dynamic}, \emph{guided} and \emph{runtime}; for details, see the Cython online documentation at \url{http://cython.readthedocs.io}).

\subsection{Calculation of the evidence value}\label{subsec:evidence}

After obtaining the MAP estimate for the segmentation point, the algorithm divides the signal in two segments, and calculates the evidence for the hypothesis $H_0:\delta =1$ that the segments have equal variance. 

This calculation is performed by the function \textit{tester} of the SeqSeg class. It starts by calculating $p_0$, the maximum posterior under $H_0$, which is done analitically. Then it runs the MCMC algorithm to calculate the posterior integral over the surprise set $T(p_0) = \{(d,s) \in \Re_+ \times \Re_+ : P(d,s|y^1, y^2) > p_0$.

In our implementation, the method that runs the MCMC is defined as both \textit{cdef} and \textit{nogil}, to achieve maximum performance and also to be possibly executed in parallel. This is the \textit{cmcmc} method in our code.

Our goal with the MCMC is to sample from the full posterior \ref{eq:fullpost} with support over $\Re_+ \times \Re_+$. To obtain the samples, we adopt the Adaptive Metropolis algorithm of Haario \cite{Haario2001}.

The simulation occurs in three phases: first, we generate $t_0$ candidate points $(d_{cand}, s_{cand})$ from candidate distributions $d_{cand} \sim \mathcal{N}(d_{cur}, \sigma_d^2)$ and $s_{cand} \sim \mathcal{N}(s_{cur}, \sigma_s^2)$ where $d_{cur}$ and $s_{cur}$ are the current points. Each point is accepted or rejected following the usual Metropolis-Hastings random walk acceptance probability. To start this phase, we initalize the state of the chain using Gaussians with the following averages

\begin{align}
 \tilde{s} = & \frac{\sum_{i=1}^{N_1} y_{1,i}^2}{N_1-1} \\
 \tilde{d} = & \frac{\sum_{i=1}^{N_2} y_{2,i}^2}{N_2 - 1} \cdot \frac{N_1-1}{\sum_{i=1}^{N_1} y_{1,i}^2}
\end{align}

where $y_{1,\cdot}$ and $N_1$ ($y_{2,\cdot}$ and $N_2$) are the points and the size of the first (second) segment. The dispersion of these initial Gaussians are set to $\tilde{s} / 3$ and $\tilde{d} / 3$.

The goal of this phase is to obtain the first estimates for the candidate covariance matrix; after $t_0$ rounds, we estimate $\sigma_d^2$, $\sigma_s^2$ and $cov(d, s)$ from the data as follows:

\begin{align}
 \sigma_d^2 = &\frac{\sum_{i=1}^{t_0} (d_i - \bar{d})^2}{t_0 - 1} \\
 \sigma_s^2 = &\frac{\sum_{i=1}^{t_0} (s_i - \bar{s})^2}{t_0 - 1} \\
 cov(d,s) = &\frac{\sum_{i=1}^{t_0} (d_i - \bar{d})(s_i - \bar{s})}{t_0 - 1}
\end{align}

Here $s_i$ and $d_i$ are the samples of the two parameters of the model, $\sigma_0$ and $\delta$, respectively.

After the estimation, the second phase of the simulation starts. This phase is a burn-in period (meaning that the samples generated during this phase will not be used in the estimation of the integral), during which the candidate points are generated simultaneously for both parameters, $(d_{cand}, s_{cand}) \sim \mathcal{N}\left(\left(d_{cur}, s_{cur}\right), \Sigma_t\right)$.

During this phase, the covariance matrix $\Sigma_t$ is updated recursively, following the method from \cite{Haario2001}:

\begin{equation}
 \Sigma_{t+1} = \frac{t-1}{t}\Sigma_t + \frac{s_d}{t}\left(t \bar{X}_{t-1}\bar{X}_{t-1}^T - (t+1)\bar{X}_t\bar{X}_t^T + \epsilon I_2\right)
\end{equation}

Here $\bar{X}_t$ represents the column vector of the sample points average up to point $t$, and $I_2$ is the 2-dimensional identity matrix.

This method needs two parameters, $s_d$ and $\epsilon$. The first is a scaling parameter which must be set according to the dimension of the parametric space we are sampling from. Following the recomendation from \cite{Haario2001} (which in turn is following an analysis by \cite{Gelman1996}), we define it to be $s_d = (2.24)^2/2$. The second parameter is present only to avoid singularity of the covariance matrix. Our analysis indicate that it must be set to a very low value; in our code, we define $\epsilon = 1e^{-30}$ (it must be kept strictly positive to guarantee ergodicity of the adaptive chain). 

After the burn-in, the actual sampling starts. In this phase, we generate candidates $(d_{cand}, s_{cand})$ from a Gaussian $\mathcal{N}\left(\left(d_{cur}, s_{cur}\right), \Sigma\right)$, where the covariance matrix is now fixed. At each step, we obtain $(d_{cur}, s_{cur})$ by applying the usual Metropolis-Hasting acceptance probabilities on the candidate points. After that, we obtain the posterior value $P(d_{cur}, s_{cur} | y^1, y^2)$, compare it to the maximum posterior under $H_0$, and keep track of the number of points with posterior greater than $p_0$. This number, divided by the chain's total length (considering only the last phase), is the evidence value $\overline{ev(H_0)}$ \textbf{against} $H_0$.

The \textit{tester} method starts \textit{nchains} parallel chains; this feature is useful if one is interested in keeping track of the chain's convergence, by using for instance Gelman-Rubin-Brooks statistics \cite{Gelman1998}. For each chain, the \textit{cmcmc} method returns the evidence value \textbf{for} $H_0$ ($1 - \overline{ev(H_0)}$); \textit{tester} then averages this evidence, and compares it to the parameter $\alpha$. If the (average) evidence for $H_0$ is greater than $\alpha$, the algorithm stops, declaring that the two segments are equivalent; otherwise, it inserts the current change point in the list, and tries to segment again each of the new segments.

The algorithm is summarized in the pseudocode \ref{algo}.

\begin{algorithm}[t]
 \caption{Sequential Segmentation Algorithm (SeqSeg)}
 \label{algo}
  \begin{algorithmic}[1]
    \Procedure{SeqSeg}{$y$, \textit{minlength}, $\alpha$}
    \State $N = dim(y)$
    \State $\bar{t} = argmax_t P(t | y)$ \Comment{Optimize the posterior for $t$}
    \If{$\bar{t} < $ \textit{minlength} \textbf{or} $N - \bar{t} < $ \textit{minlength}} 
      \State \textbf{return}
    \Else{}
    \State $y^1 = \{y_i : i < \bar{t}\}$, $y^2 = \{y_i : i \geq \bar{t}\}$
    \State $p0 = max_{\delta = 1} P(\sigma | y^1. y^2)$ \Comment{Max. posterior under $H_0$}
    \State $ev = \int_{T_{p0}}P(\sigma, \delta | y^1, y^2)$ \Comment{Evidence against $H_0$}
    \If{$1-ev < \alpha$} 
      \State $t_1 =$ \textbf{SeqSeg}($y^1$, \textit{minlength}, $\alpha$)
      \State $t_2 =$ \textbf{SeqSeg}($y^2$, \textit{minlength}, $\alpha$)
      \State \textbf{return} $[t_1, \bar{t}, \bar{t} + t_2]$ 
    \Else{}
      \State \textbf{return}
    \EndIf
    \EndIf
    \EndProcedure
  \end{algorithmic}
\end{algorithm}

\section{Sensibility to parameters}\label{sec:param}

\subsection{Time resolution for cutpoint estimation}\label{subsec:tres}

To appropriately apply the SeqSeg algorithm, a few choices must be made. First, one has to pick the minimum segment length, and the time resolution. The minimum segment length can be set arbitrarily; in most cases the algorithm will stop before reaching small segments, and this parameter has little or no effect. We recommend setting it to a size representing what one expects from the data; in subacquatic audio signals, for instance, we usually avoid segments of less than half a second, since these would hardly contain a significant or interesting event.

The time resolution, on the other hand, have a direct impact on the performance of the algorithm, since it dictates the number of function evaluations during the change point estimation step. However, since the \textit{bayeseg} module implements a parallel version of this step, and all the calculations involved are fully optimized, it is possible to achieve good performance with a resolution close or equal to 1. Of course, the impact of the parallelization is strongly dependent on the configuration of the system that will run the algorithm. 

In table \ref{tbl:table1}, we report computing time for the segmentation step only, on a system with 4 $\times$ 1.6 GHz i5 processors and 8 Gb RAM memory, running Ubuntu 16.04. We simulate signals with lengths of $10,000$, $100,000$ and $1,000,000$ points, and use time resolutions of $1$, $10$, $100$ and $1,000$. The signals are simulated from a Gaussian distribution with $0$ mean and standard deviation of $1$.

\begin{table}
\centering
\caption{Effect of time resolution on running time}
\label{tbl:table1}
\begin{tabular}{l|l|l}
Signal size & Resolution & Time (s) \\ \hline
10000 & 1 & 0.01982 \\
10000 & 10 & 0.009003 \\
10000 & 100 & 0.008039 \\
10000 & 1000 & 0.004694 \\
100000 & 1 & 0.05072 \\
100000 & 10 & 0.007656 \\
100000 & 100 & 0.002263 \\
100000 & 1000 & 0.001422 \\
1000000 & 1 & 0.6214 \\
1000000 & 10 & 0.1034 \\
1000000 & 100 & 0.02704 \\
1000000 & 1000 & 0.02186 \\
\end{tabular}
\end{table}

As we see, the total elapsed time necessary to evaluate the posterior over all points in a $1,000,000$ long signal is in the order of $0.6$ seconds; this means that, when running on a system with many cores, one can safely set the time resolution to 1 and guarantee maximum precision of the algorithm, without incurring in large processing costs.

\subsection{Convergence of MCMC step}\label{subsec:mcmc}

After defining the minimum segment length and the time resolution, the next step should be to define the sample sizes for the MCMC. This can be done by using simulated or real signals, by starting the algorithm with small values for \textit{mcburn} and \textit{mciter}, running the segmentation multiple times and observing the variance of the number of segments found. If this number variates between multiple runs of the segmentation, it is an indication that the chain must be simulated for longer times. This is the case because our algorithm is completely deterministic, except for the calculation of the evidence value. This calculation, however, should converge whenever the MCMC method converge sufficiently itself.

It is also possible (and advisable) to run multiple parallel chains and monitor convergence measures such as the Gelman-Rubin $\hat{R}$ statistic \cite{Gelman1998}. This statistic is obtained in the following way: suppose we run $M$ independent chains, starting from different points (drawn from an overdispersed distribution). From each chain we obtain $n$ samples of the parameter $\theta$ of interest, and define 

\begin{align}
 \hat{\theta}_m = & \frac{1}{n}\sum_{i=1}^n \theta_{m,i} \\
 \hat{\sigma}^2_m = & \frac{1}{n-1}\sum_{i=1}^n (\theta_{m,i} - \hat{\theta}_m)^2 \\
 \hat{\theta} = & \frac{1}{M}\sum_{i=1}^M \hat{\theta}_m \\
 B = & \frac{n}{M-1}\sum_{i=1}^M \left(\hat{\theta}_m - \hat{\theta}\right)^2 \\
 W = & \frac{1}{M}\sum_{i=1}^M \hat{\sigma}^2_m 
\end{align}

Finally, we define the pooled variance $\hat{V}$ as 

\begin{equation}
 \hat{V} = \frac{n-1}{M}W + \frac{M+1}{Mn}B
\end{equation}

and the Gelman-Rubin statistic $\hat{R}$ as 

\begin{equation}
 \hat{R} = \frac{\hat{V}}{W}
\end{equation}

The value of $\hat{R}$ should approach 1 for appropriately convergent chains. If it is higher than 1, longer samples are needed to approach the posterior distribution.

In our algorithm, the MCMC chains will be used to test equality of variances between two segments. To analyze the test's behavior we simulate gaussian signals with $1,000,000$ points, with standard deviations $1$, and $\delta \in \{1, 1.1, 1.5, 2\}$. When $\delta > 1$, the cutting point is at $\bar{t} = 500,000$.

We use chains of length $\{1000, 10000, 100000\}$, and run $M = 4$ parallel chains in each situation. For all simulations, we adopt $\beta = 1$. In table \ref{tbl:table2} we report the $\hat{R}$ statistic, and other summary measures for the chains.

\begin{sidewaystable}
\centering
\caption{Analysis of the FBST results}
\label{tbl:table2}
\begin{tabular}{l|r|r|r|r|r|r|r|r|r}
 $\delta$ &  MCMC sample size & Minimum evidence & Maximum evidence & Acceptance rate & $\hat{\delta}$ & $\hat{\sigma}_0$ & $\hat{R}_\delta$ &  $\hat{R}_\sigma$ \\ \hline

1.0 &    1000 &  1.00000 &  1.00000 &  0.019250 &  1.000802 &  0.998831 &  1.239326 &  1.097055 \\
1.0 &   10000 &  0.99760 &  1.00000 &  0.061075 &  0.999650 &  0.998841 &  1.006411 &  1.001696 \\
1.0 &  100000 &  0.99562 &  0.99747 &  0.147852 &  0.999800 &  0.998811 &  1.000015 &  1.000212 \\
1.1 &    1000 &  0.00000 &  0.00000 &  0.014500 &  1.103171 &  0.998679 &  1.081629 &  1.053349 \\
1.1 &   10000 &  0.00000 &  0.00000 &  0.088725 &  1.102898 &  0.998835 &  1.005252 &  1.003423 \\
1.1 &  100000 &  0.00000 &  0.00000 &  0.126205 &  1.102852 &  0.998865 &  1.000383 &  1.000240 \\
1.5 &    1000 &  0.00000 &  0.00000 &  0.034500 &  1.503393 &  0.998099 &  1.179845 &  1.041876 \\
1.5 &   10000 &  0.00000 &  0.00000 &  0.089450 &  1.502498 &  0.998415 &  1.002443 &  1.002263 \\
1.5 &  100000 &  0.00000 &  0.00000 &  0.126858 &  1.502448 &  0.998406 &  1.000255 &  1.000358 \\

\end{tabular}
\end{sidewaystable}

We see from table \ref{tbl:table2} that a sample size of around $10,000$ shows good convergence measures already. The minimum and maximum evidence are very tight even for small sample sizes, which is an effect of the small posterior variance. The stopping criteria based on the FBST would be effective in all of these situations, segmentating the signals when $\delta > 1$, and keeping the original signal when $\delta = 1$.

It is important to note a few things about our algorihtm and the characteristics of the signals we want to segmentate. These are audio files with a duration of 15 minutes, at a sampling rate of $f_s = 11,025$. This means that our original signal will be of size $N = f_s \cdot T = 9,922,500$. Supposing for simplicity that the segmentation point is exactly at the middle of the signal, this gives two segments of roughly $4,500,000$ points, whose variances we need to compare. 

With a sample size of this order, and assuming our model is correct, the posterior will be extremely concentrated around its mode; so care must be taken when running the chains, to avoid them to have too small an acceptance ratio. This is accomplished by the above mentioned adaptive algorithm, that keeps track of the covariance matrix of the Gaussian candidate.

\subsection{Prior for $\delta$ and decision function threshold}\label{subsec:prior}

Now there remains the $\beta$ and $\alpha$ parameters. These two parameters control the algorithm's sensibility: lower values of $\beta$ concentrate the prior distribution of $\delta$ around $\delta = 1$, thus demanding more evidence from the data to accept different values of this parameter. Lower values of $\alpha$ make the algorithm accept equality of variances for lower values for the evidence, thus making the algorithm less prone to accept segments. 

To evaluate the performance of the algorithm for different values of $\alpha$ and $\beta$, we simulate a signal with a total length of $1,000,000$ points; we use a Gaussian distribution with 0 mean and standard deviation of 1. We then change the signal's variance in different segments, by multiplying the simulated values by $\delta$; we simulate a total of 6 segments, with boundaries given by $\{$ $0$; $10,000$; $110,000$; $200,000$; $500,000$; $750,000$; $1,000,000$ $\}$. We create alternated segments, the first always having standard deviation of $1$. 

For each test $\delta$ is fixed; we run three batches of tests with $\delta = 1$ in the first, $\delta = 1.1$ in the second, and finally $\delta = 1.5$. 

Finally, for each simulated signal with a given value of $\delta$, we run the segmentation algorithm 30 times for each combination of $\beta \in \{$ $1$, $0.1$, $0.01$, $0.001$, $0.0001$, $0.00001$ $\}$ and $\alpha \in \{0.1, 0.5, 0.9, 0.99\}$. We compute the average elapsed time for each combination of values, and also the minimum and maximum of the number of segments returned by the algorithm. Tables \ref{tbl:table3} through \ref{tbl:table6} present the results.

\begin{table*}
\centering
\caption{Algorithm results for $\alpha = 0.1$}
\label{tbl:table3}
\begin{tabular}{c|c|c|c|c}
$\delta$ &     $\beta$ &  Min \# of segments & Max \# of segments & Average time (s) \\ \hline
   1.0 &  1.00000 &       1 &       1 &  0.013725 \\
   1.0 &  0.10000 &       1 &       1 &  0.013490 \\
   1.0 &  0.01000 &       1 &       1 &  0.013503 \\
   1.0 &  0.00100 &       1 &       1 &  0.013505 \\
   1.0 &  0.00010 &       1 &       1 &  0.016907 \\
   1.0 &  0.00001 &       1 &       1 &  0.018005 \\
   1.1 &  1.00000 &       6 &       6 &  0.098041 \\
   1.1 &  0.10000 &       6 &       6 &  0.113072 \\
   1.1 &  0.01000 &       6 &       6 &  0.096792 \\
   1.1 &  0.00100 &       5 &       5 &  0.097185 \\
   1.1 &  0.00010 &       1 &       1 &  0.040666 \\
   1.1 &  0.00001 &       1 &       1 &  0.029612 \\
   1.5 &  1.00000 &       6 &       6 &  0.093532 \\
   1.5 &  0.10000 &       6 &       6 &  0.097517 \\
   1.5 &  0.01000 &       6 &       6 &  0.096968 \\
   1.5 &  0.00100 &       6 &       6 &  0.097118 \\
   1.5 &  0.00010 &       3 &       4 &  0.064936 \\
   1.5 &  0.00001 &       1 &       1 &  0.028431 \\

\end{tabular}
\end{table*}

\begin{table*}
\centering
\caption{Algorithm results for $\alpha = 0.5$}
\label{tbl:table4}
\begin{tabular}{c|c|c|c|c}
$\delta$ &     $\beta$ &  Min \# of segments & Max \# of segments & Average time (s) \\ \hline
   1.0 &  1.00000 &       1 &       1 &  0.013530 \\
   1.0 &  0.10000 &       1 &       1 &  0.013495 \\
   1.0 &  0.01000 &       1 &       1 &  0.013553 \\
   1.0 &  0.00100 &       1 &       1 &  0.013726 \\
   1.0 &  0.00010 &       1 &       1 &  0.013643 \\
   1.0 &  0.00001 &       1 &       1 &  0.018850 \\
   1.1 &  1.00000 &       6 &       6 &  0.096758 \\
   1.1 &  0.10000 &       6 &       6 &  0.096947 \\
   1.1 &  0.01000 &       6 &       6 &  0.097188 \\
   1.1 &  0.00100 &       5 &       5 &  0.096727 \\
   1.1 &  0.00010 &       1 &       1 &  0.028671 \\
   1.1 &  0.00001 &       1 &       1 &  0.029201 \\
   1.5 &  1.00000 &       6 &       6 &  0.093676 \\
   1.5 &  0.10000 &       6 &       6 &  0.100382 \\
   1.5 &  0.01000 &       6 &       6 &  0.096994 \\
   1.5 &  0.00100 &       6 &       6 &  0.096927 \\
   1.5 &  0.00010 &       4 &       4 &  0.083853 \\
   1.5 &  0.00001 &       1 &       1 &  0.028569 \\
\end{tabular}
\end{table*}

\begin{table*}
\centering
\caption{Algorithm results for $\alpha = 0.9$}
\label{tbl:table5}
\begin{tabular}{c|c|c|c|c}
$\delta$ &     $\beta$ &  Min \# of segments & Max \# of segments & Average time (s) \\ \hline
   1.0 &  1.00000 &       1 &       1 &  0.015604 \\
   1.0 &  0.10000 &       1 &       1 &  0.013489 \\
   1.0 &  0.01000 &       1 &       1 &  0.013576 \\
   1.0 &  0.00100 &       1 &       1 &  0.013651 \\
   1.0 &  0.00010 &       1 &       1 &  0.013650 \\
   1.0 &  0.00001 &       1 &       1 &  0.013503 \\
   1.1 &  1.00000 &       6 &       6 &  0.093792 \\
   1.1 &  0.10000 &       6 &       6 &  0.097041 \\
   1.1 &  0.01000 &       6 &       6 &  0.096902 \\
   1.1 &  0.00100 &       5 &       5 &  0.098112 \\
   1.1 &  0.00010 &       1 &       1 &  0.028455 \\
   1.1 &  0.00001 &       1 &       1 &  0.028557 \\
   1.5 &  1.00000 &       6 &       6 &  0.093640 \\
   1.5 &  0.10000 &       6 &       6 &  0.098186 \\
   1.5 &  0.01000 &       6 &       6 &  0.097044 \\
   1.5 &  0.00100 &       6 &       6 &  0.097160 \\
   1.5 &  0.00010 &       4 &       4 &  0.083806 \\
   1.5 &  0.00001 &       1 &       1 &  0.028375 \\
\end{tabular}
\end{table*}

\begin{table*}
\centering
\caption{Algorithm results for $\alpha = 0.99$}
\label{tbl:table6}
\begin{tabular}{c|c|c|c|c}
$\delta$ &     $\beta$ &  Min \# of segments & Max \# of segments & Average time (s) \\ \hline
   1.0 &  1.00000 &       1 &       1 &  0.013522 \\
   1.0 &  0.10000 &       1 &       1 &  0.013504 \\
   1.0 &  0.01000 &       1 &       1 &  0.013584 \\
   1.0 &  0.00100 &       1 &       1 &  0.013589 \\
   1.0 &  0.00010 &       1 &       1 &  0.013940 \\
   1.0 &  0.00001 &       1 &       1 &  0.013625 \\
   1.1 &  1.00000 &       6 &       6 &  0.103654 \\
   1.1 &  0.10000 &       6 &       6 &  0.096897 \\
   1.1 &  0.01000 &       6 &       6 &  0.097490 \\
   1.1 &  0.00100 &       5 &       5 &  0.097019 \\
   1.1 &  0.00010 &       1 &       1 &  0.028837 \\
   1.1 &  0.00001 &       1 &       1 &  0.028453 \\
   1.5 &  1.00000 &       6 &       6 &  0.093719 \\
   1.5 &  0.10000 &       6 &       6 &  0.101384 \\
   1.5 &  0.01000 &       6 &       6 &  0.096989 \\
   1.5 &  0.00100 &       6 &       6 &  0.097059 \\
   1.5 &  0.00010 &       4 &       4 &  0.082309 \\
   1.5 &  0.00001 &       1 &       1 &  0.039567 \\
\end{tabular}
\end{table*}

For all these tests we used MCMC samples of size $10,000$ after burning another $10,000$ points. Whenever the MCMC step is sufficiently long, we do not expect variations between runs of the algorithm over the same signal. This is indeed what we see in the results, except for $\alpha = 0.1$, when there is a small variation in the number of segments for $\delta = 1.5$ and $\beta = 0.0001$.

The algorithm results were the same for all different values of $\alpha$. This means that, in these tests, the evidence value was very close to the extremes of $0$ and $1$ when the segmentation point correctly identified a segment (in the first case) or when the segmentation did not capture a change in variance (in the second case). 

These tests indicate that the algorithm is very robust, specially to values of $\alpha$, but also for values of $\beta$ in a wide range. Values of $\beta$ that are too small lead to undersegmentation of the signal (i.e., the algorithm ignores some existent segmentation points), as we expected; but when $\delta = 1.1$, the algorithm's output was the same for $\beta \in \{0.01, 0.1, 1\}$; when $\delta = 1.5$, i.e., when the difference in variance is higher, it identified the correct number of segments for $\beta \in \{0.001, 0.01, 0.1, 1\}$.

Now there remains the test for different $\delta$ values in each segment. To simulate this situation, we keep the same structure as above ($1,000,000$ points with segment boundaries at $\bar{t} \in \{$ $0$; $10,000$; $110,000$; $200,000$; $500,000$; $750,000$; $1,000,000$ $\}$) but now we use $\delta = 1.1$ for the second segment, $\delta = 1.5$ for the fourth segment, and $\delta = 1.2$ for the sixth. We fix $\alpha = 0.1$ and run the algorithm for $\beta$ in an evenly spaced grid of $100$ points between $1e^{-5}$ and $1e^{-3}$, then again on another grid of $100$ points between $1e^{-3}$ and $1e^{-1}$.

\begin{figure}[H]
   \centering
   \includegraphics[width=.4\textwidth]{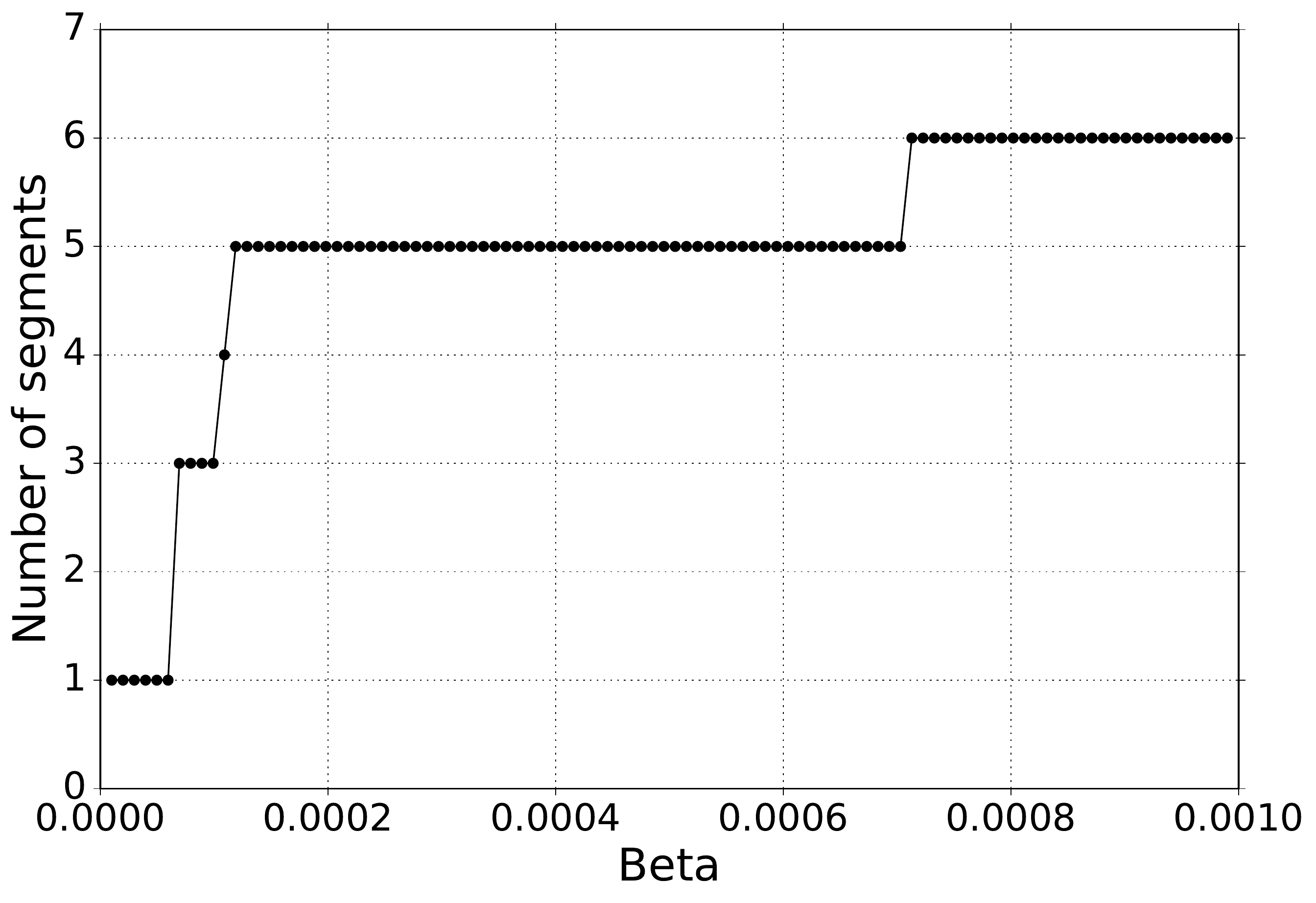}
   \caption{Number of segments for different values of $\beta$ ($\alpha = 0.1$)}
   \label{fig:beta1}
\end{figure}

\begin{figure}[H]
   \centering
   \includegraphics[width=.4\textwidth]{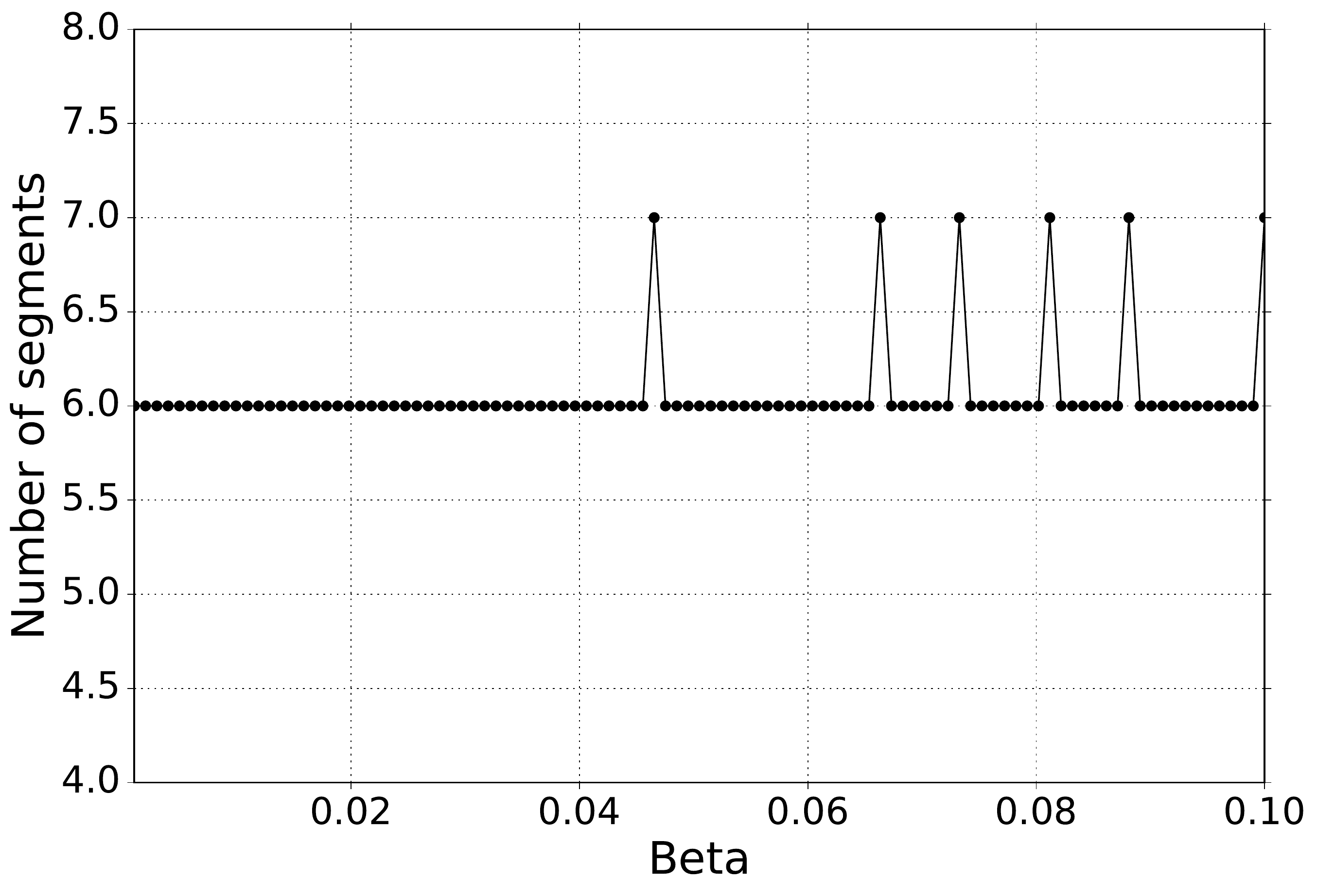}
   \caption{Number of segments for different values of $\beta$ ($\alpha = 0.1$)}
   \label{fig:beta2}
\end{figure}

The results show that the number of segments is 1 when $\beta$ is very low (less than $4e^{-05}$). After that, the number of segments grows quickly, reaching $5$ segments when $\beta \approx 0.0001$. After this threshold, the algorithm becomes more stable; finally, it finds the correct number of segments when $\beta \approx 0.0007$, and then the behavior completely stabilizes apart from a few values of $\beta$ for which there is one more segment. This behavior is illustrated in figures \ref{fig:beta1} and \ref{fig:beta2}.

Recalling the ground truth segments

\begin{enumerate}
 \item From $i = 0$ to $i = 10,000$, with variance $1$;
 \item From $i = 10,001$ to $i = 110,000$, with variance $1.1$;
 \item From $i = 110,001$ to $i = 200,000$, with variance $1$;
 \item From $i = 200,001$ to $i = 500,000$, with variance $1.5$;
 \item From $i = 500,001$ to $i = 750,000$, with variance $1$;
 \item From $i = 750,001$ to $i = 1,000,000$, with variance $1.2$
\end{enumerate}

we see that the algorithm first divided the signal in order to completely separate the higher variance segment ($500,000 < i < 750,000$). This also separates the segment with the second highest variance ($750,000 < i < 1,000,000$). After that, the change points at $i = 200,000$ and $i = 110,000$ are found; the last segment to be identified is the smallest one, at $i = 10,000$.

Also we note that, when there is any oversegmentation (for some higher values of $\beta$), it always occurs inside the segment from $i = 500,000$ to $i = 750,000$.

\begin{figure}[H]
   \centering
   \includegraphics[width=.4\textwidth]{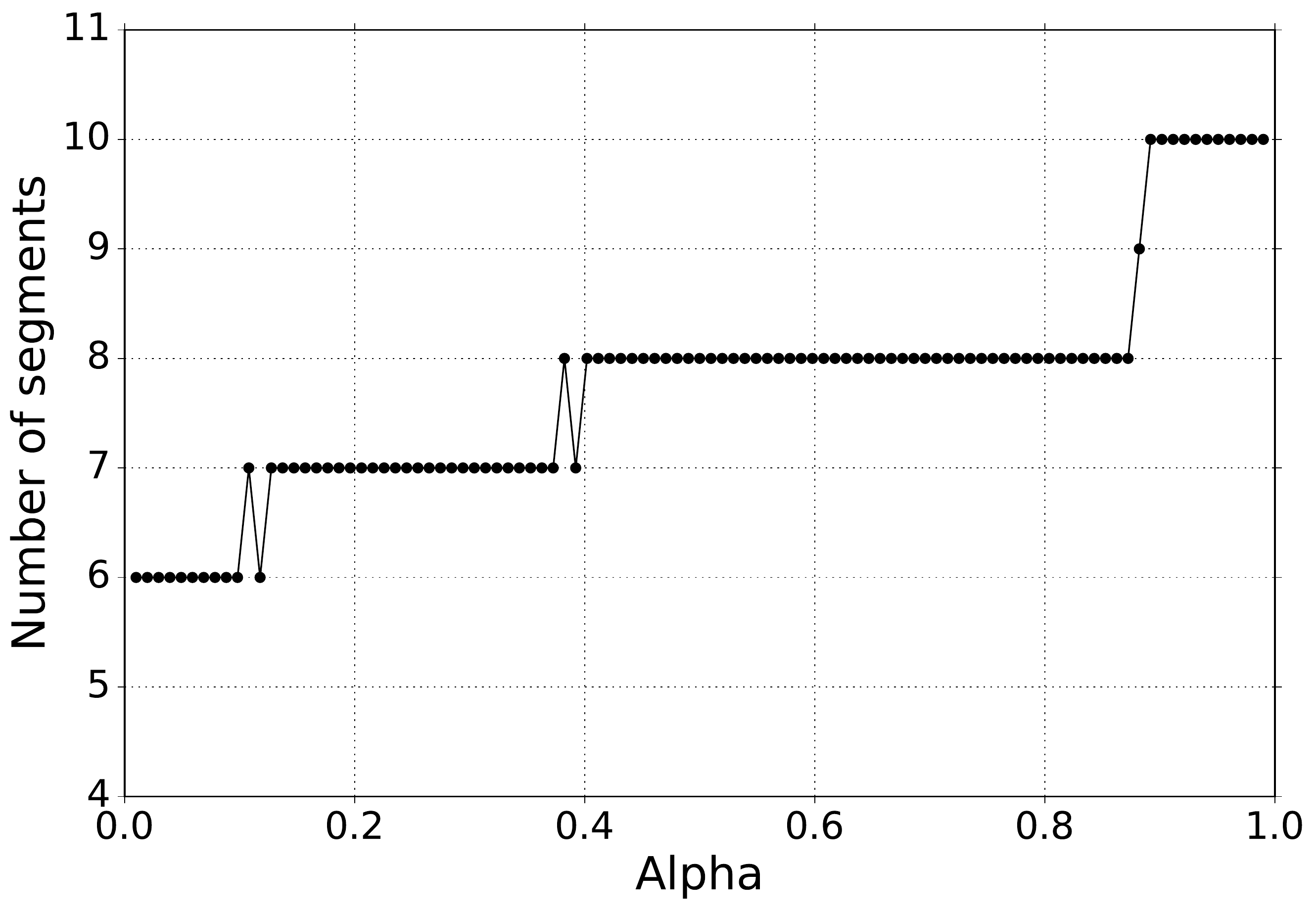}
   \caption{Number of segments for different values of $\alpha$ ($\beta = 0.1$)}
   \label{fig:alpha}
\end{figure}

Now, if we fix $\beta = 0.001$ and use a $100$ points uniform grid for $\alpha$ between $[0.01, 0.99]$, we find that the number of segments is the same for all values of $\alpha$ in this situation, indicating again that the algorithm is very robust to choices of this parameter whenever $\beta$ is well-calibrated. However, if we fix $\beta = 0.1$ and vary $\alpha$ in the same grid, we see oversegmentation occuring for higher values of $\alpha$, see figure \ref{fig:alpha}.

From the results of this section, then, we conclude that the critical parameter to be calibrated is $\beta$; when it is well-calibrated, the algorithm becomes unsensitive to changes in $\alpha$. When $\beta$ is too high, the algorithm will tend to oversegmentate the signal; this oversegmentation will grow with $\alpha$.

These observations indicate that, when calibrating the algorithm on real signals, the best procedure is to use low values for $\alpha$ ($\alpha < 0.1$), and find the $\beta$ value that stabilizes the number of segments as equal to the ground truth. In the next section, we apply this procedure in the calibration of the parameters using real samples from the OceanPod.

\section{Application: event detection on subacquatic signals}\label{sec:app}

In this section, we apply the sequential segmentation algorithm to its original motivating task: the segmentation of subacquatic acoustic signals.

The team that operates the OceanPod, in cooperation with the administration of the \emph{Parque da Laje} administration, collected 2 years of audio recordings (2015 and 2016). These recordings are organized as 15 minutes long \emph{.wav} files with a single channel. 

TO illustrate the behavior of the algorithm, we select a few files from the collection that have distinct characteristics: one when visual inspection of the signal's spectrogram doesn't reveal any significative events; one in which we can identify one long duration event, and a third where we can identify many short duration events. Also, these samples are the same that we analyzed in \cite{Hubert2018} using the pilot version of the algorithm; we also repeat the same parameters used in this previous work, in order to be able to compare our computing times in the two versions of the algorithm.

The first file we use is a recording from 30 January 2015, Saturday, from 02:02:56 to 02:17:56. During this period, there is no perceivable activity beyond background noise (concentrated around $5$ kHz). When applying the segmentation to this sample, we would then expect no segments to be found. 

\begin{figure}[H]
   \centering
   \includegraphics[width=0.4\textwidth]{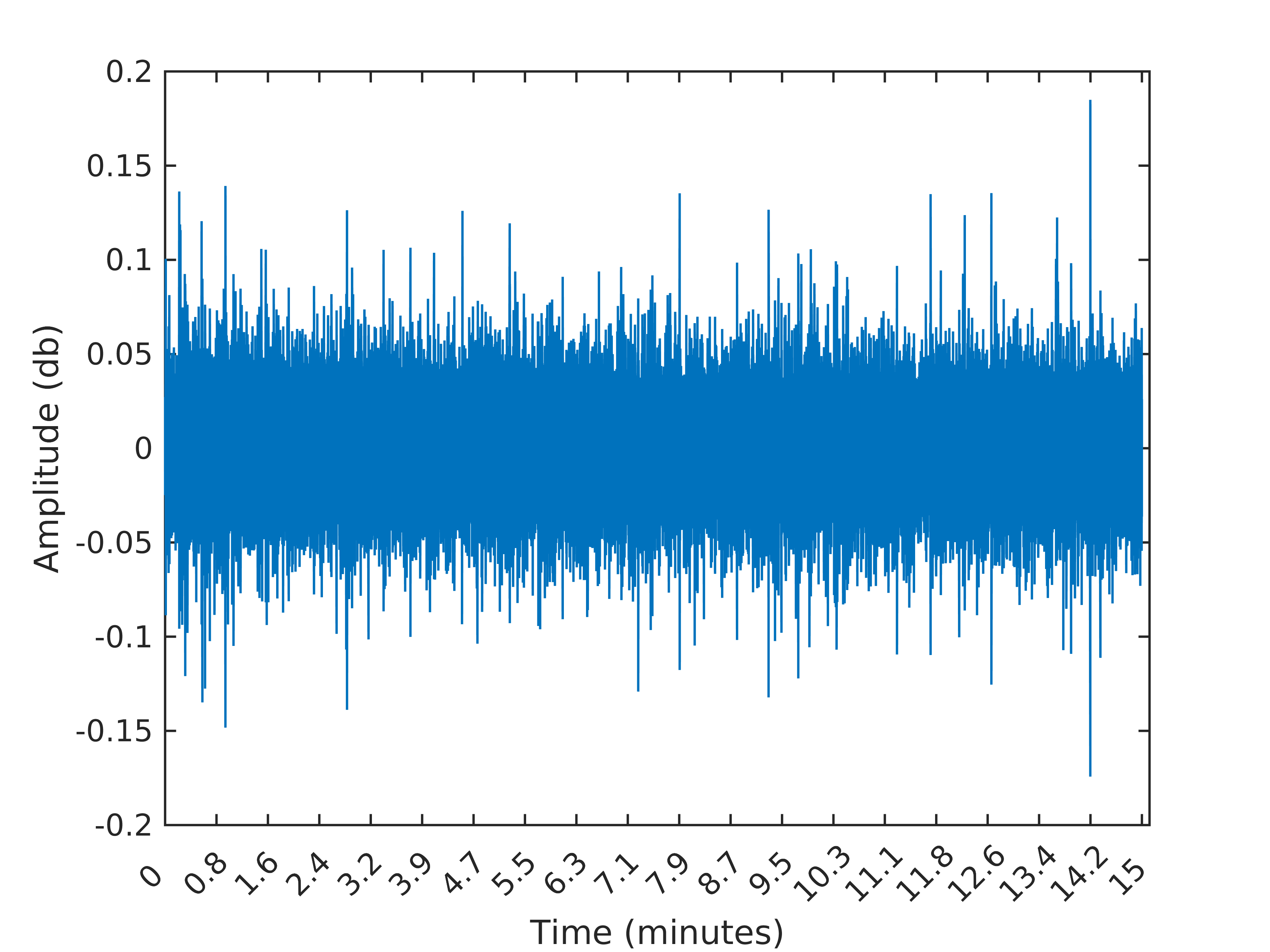}
   \caption{Waveform of the first sample: noise only}
   \label{fig:sig3101}
\end{figure}

\begin{figure}[H]
   \centering
   \includegraphics[width=0.5\textwidth]{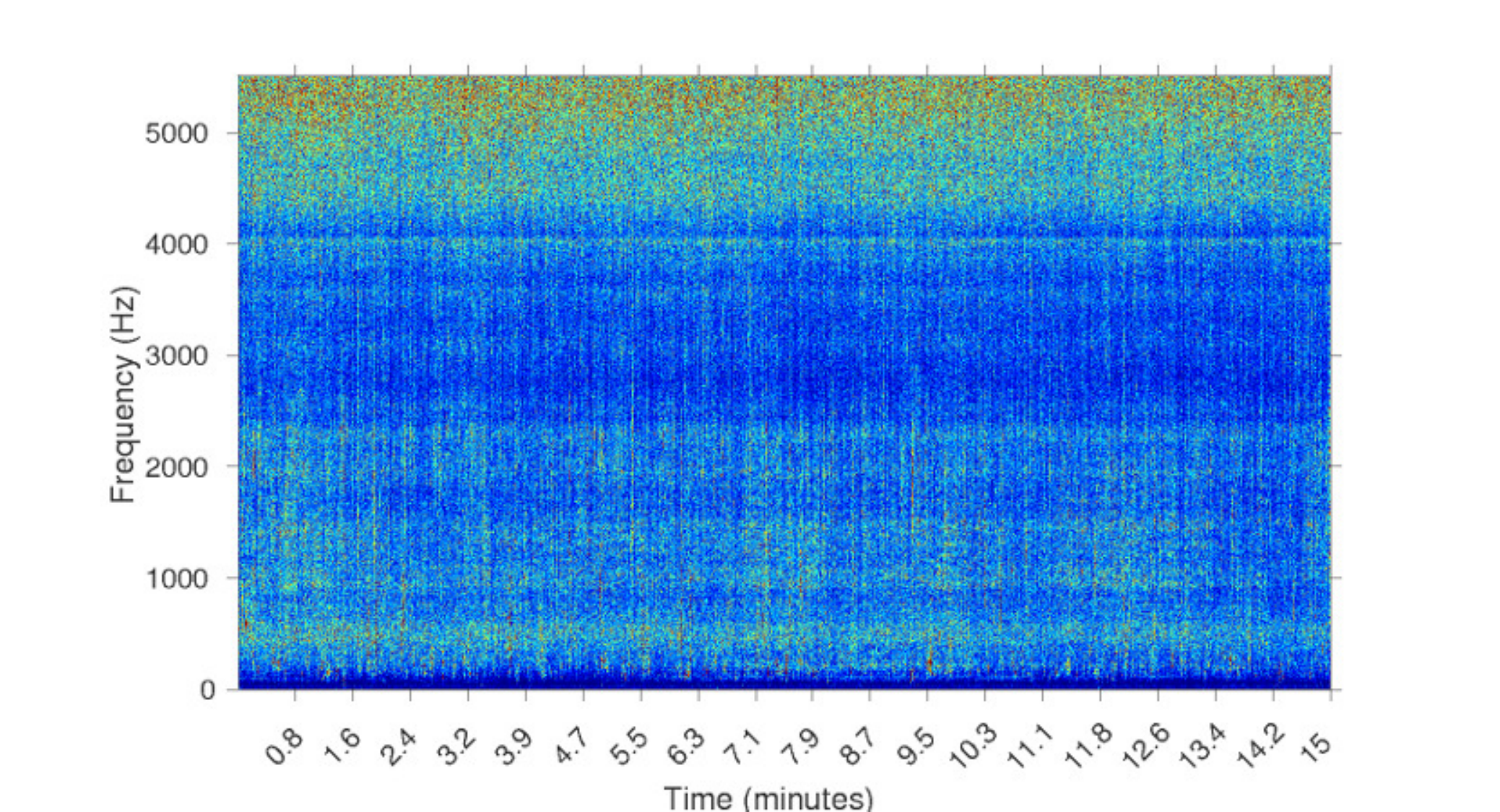}
   \caption{Spectrogram of the first sample: noise only}
   \label{fig:spec3101}
\end{figure}

The second is a recording from 2 February 2015, Monday, from 07:50:49 to 08:04:49. In this sample, we find a long duration event, starting at time $0$ and lasting for approximately $10$ min. By listening to the sample, we identify the sound of a large sized vessel, passing by at a long distance and with low speed. The segmentation algorithm applied to this sample should detect one or two change points for the signal's power, ideally forming a segment starting around the beginning of the signal and ending around $i = 6,615,000$, i.e., $10$ min into the signal (recall that the sampling rate is $f_s = 11.025$ Hz). 

\begin{figure}[H]
   \centering
   \includegraphics[width=0.4\textwidth]{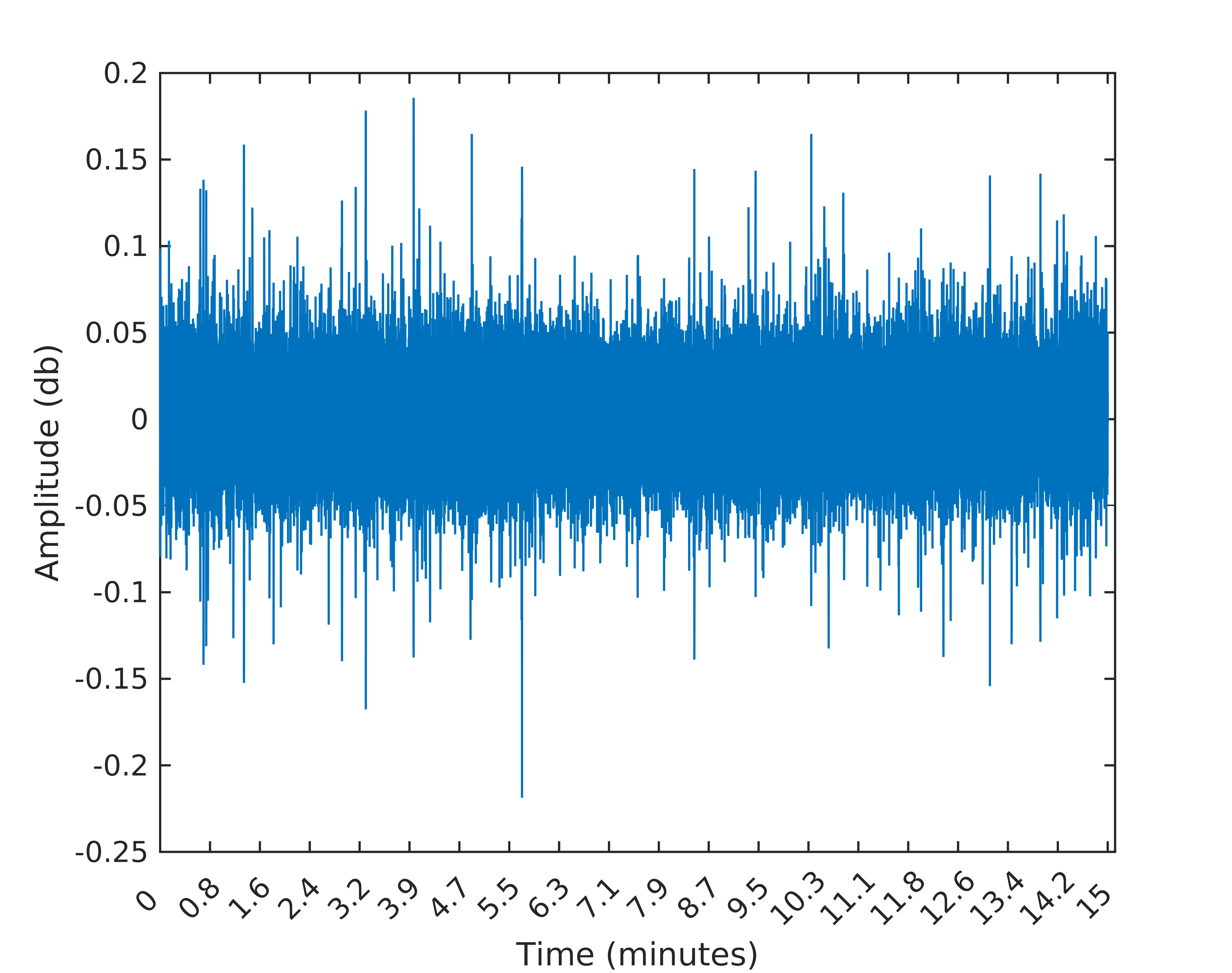}
   \caption{Waveform of the second sample: one long duration event}
   \label{fig:sig0202}   
\end{figure}

\begin{figure}[H]
   \centering
   \includegraphics[width=0.5\textwidth]{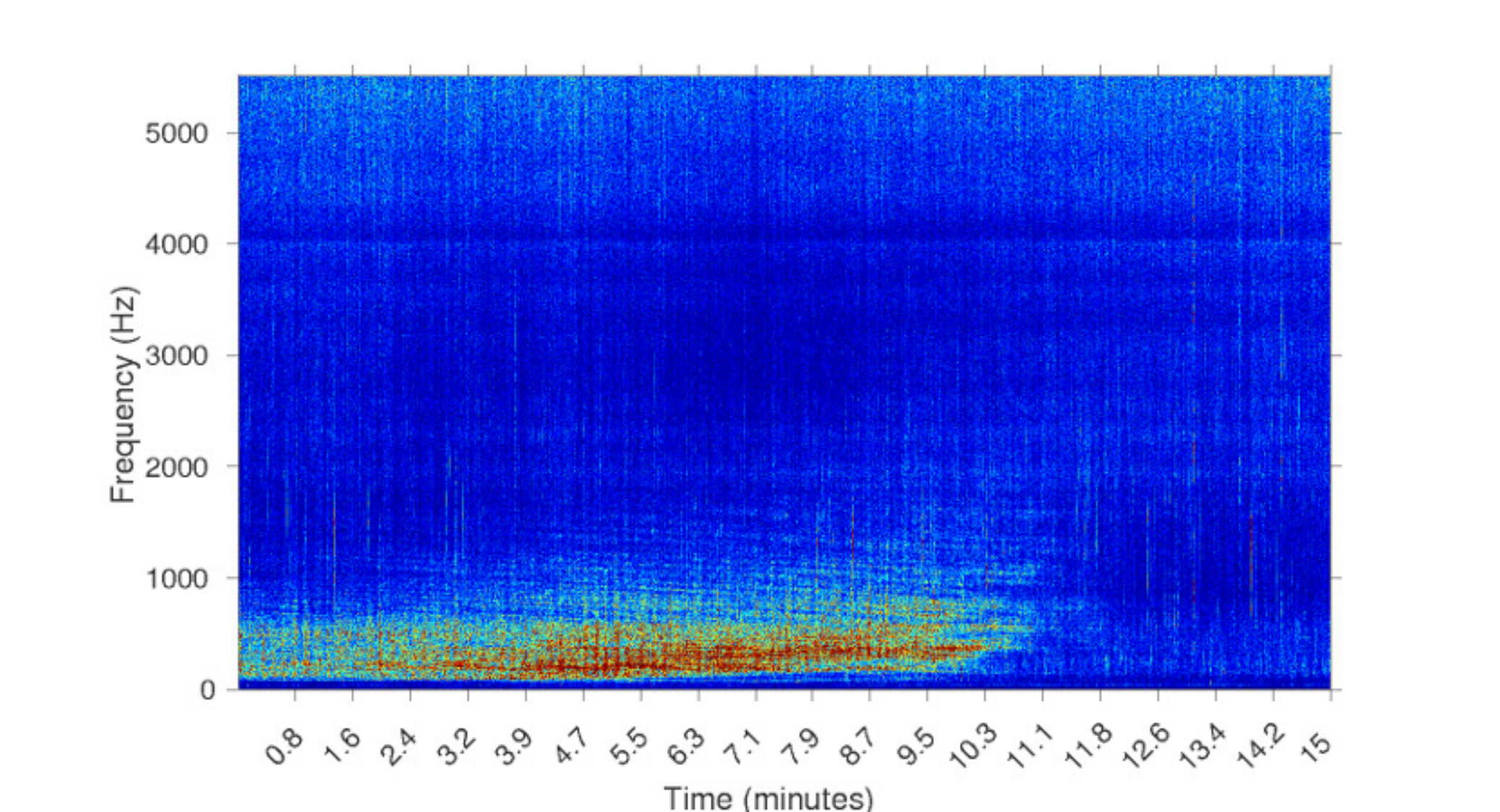}
   \caption{Spectrogram of the second sample: one long duration event}
   \label{fig:spec0202}
\end{figure}

The third sample is from 8 February 2015, Sunday, from 11:26:39 to 11:41:39. During this 15 minutes, there are many events taking place; listening to this sample, we identify the engine of smaller vessels, being turned on and off and near the hydrophone spot. In this sample, we expect the segmentation algorithm to detect many events. 

\begin{figure}[H]
   \centering
   \includegraphics[width=0.4\textwidth]{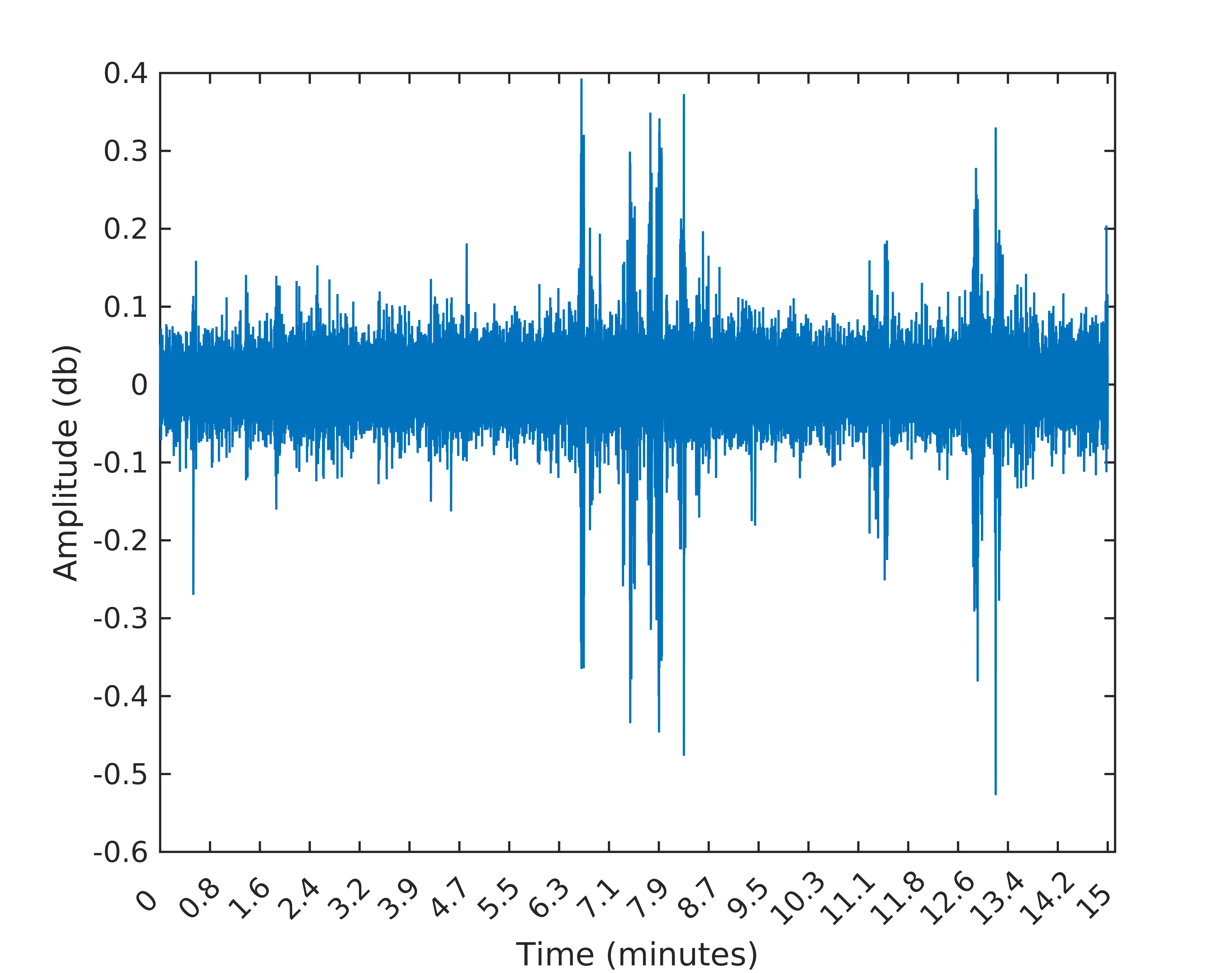}
   \caption{Waveform of the third sample: many short events}
   \label{fig:sig0802} 
\end{figure}

\begin{figure}[H]
   \centering
   \includegraphics[width=0.5\textwidth]{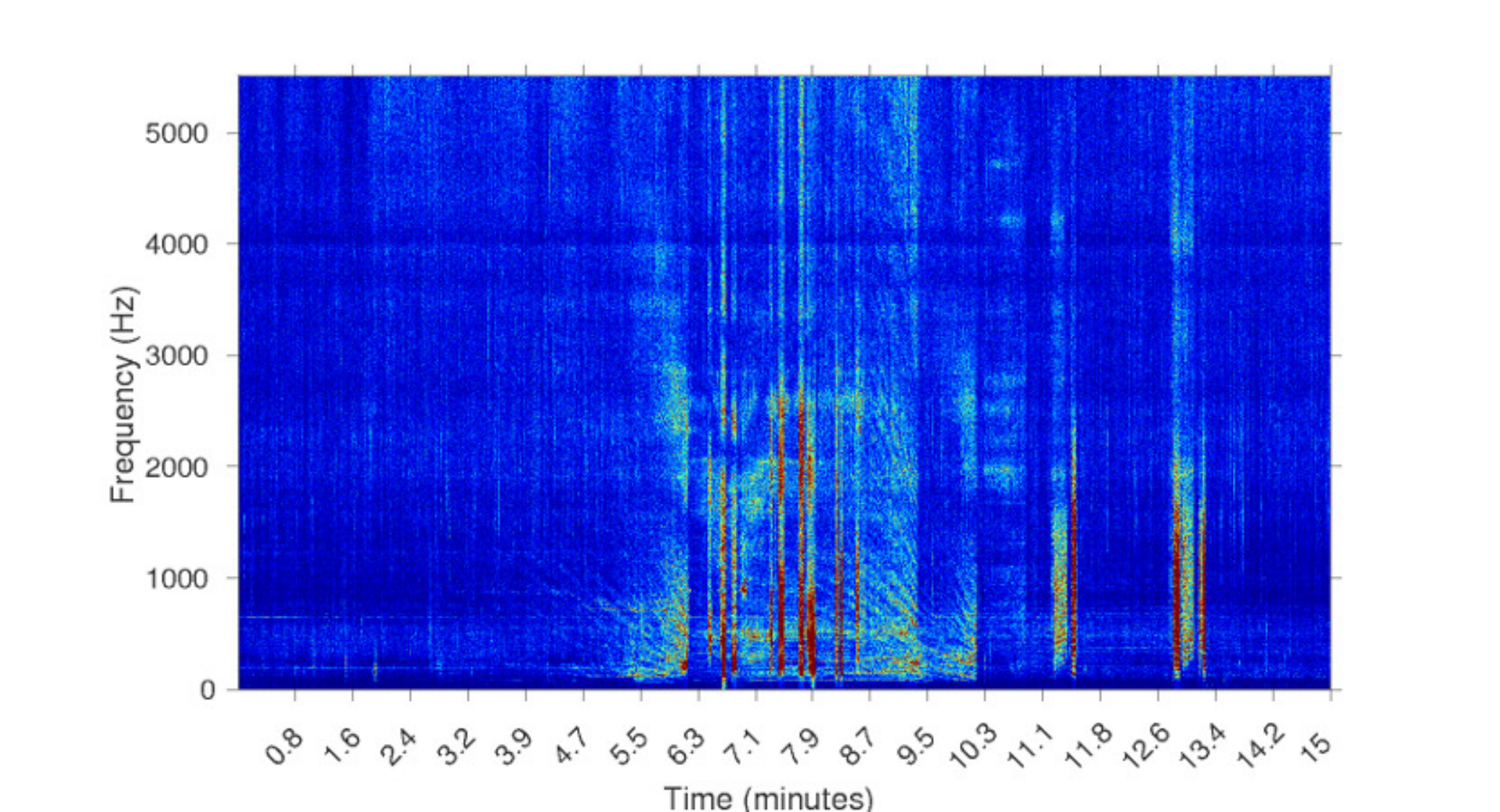}
   \caption{Spectrogram of the third sample: many short events}
   \label{fig:spec0802}
\end{figure}

As a first test, we set the mininum segment length and the time resolution to $11,025$, and obtain $20,000$ samples from the MCMC step. Following the observations from last section, we fix $\alpha = 0.1$ and run the algorithm on a grid for $\beta$. We set $\beta \in [5e-6, 5e-5]$ and use $100$ points on the grid. A summary of the results appear in table \ref{tbl:table7}. The plots of number of segments as functions of $\beta$ appear in figure \ref{fig:seg} below.

\begin{table*}
\centering
\caption{Summary of segmentation of real samples}
\label{tbl:table7}
\begin{tabular}{c|c|c|c}
Sample & $\beta$ &  \# of segments &  Average time (s) \\ \hline
 2015.01.30\_02.02.56.wav &  0.000005 &            1.0 &      0.195583 \\
 2015.01.30\_02.02.56.wav &  0.000010 &            1.0 &      0.136689 \\
 2015.01.30\_02.02.56.wav &  0.000014 &            2.0 &      0.194111 \\
 2015.01.30\_02.02.56.wav &  0.000028 &            3.0 &      0.253772 \\
 2015.01.30\_02.02.56.wav &  0.000041 &            3.0 &      0.238949 \\
 2015.01.30\_02.02.56.wav &  0.000050 &            3.0 &      0.249274 \\
 2015.02.02\_07.50.49.wav &  0.000005 &            3.0 &      0.207852 \\
 2015.02.02\_07.50.49.wav &  0.000010 &            3.0 &      0.228707 \\
 2015.02.02\_07.50.49.wav &  0.000014 &            4.0 &      0.270753 \\
 2015.02.02\_07.50.49.wav &  0.000028 &            7.0 &      0.435874 \\
 2015.02.02\_07.50.49.wav &  0.000041 &            9.0 &      0.528216 \\
 2015.02.02\_07.50.49.wav &  0.000050 &           11.0 &      0.661587 \\
 2015.02.08\_11.26.39.wav &  0.000005 &            7.0 &      0.463836 \\
 2015.02.08\_11.26.39.wav &  0.000010 &           13.0 &      0.741053 \\
 2015.02.08\_11.26.39.wav &  0.000014 &           15.0 &      1.319725 \\
 2015.02.08\_11.26.39.wav &  0.000028 &           27.0 &      1.429226 \\
 2015.02.08\_11.26.39.wav &  0.000041 &           30.0 &      1.581087 \\
 2015.02.08\_11.26.39.wav &  0.000050 &           32.0 &      1.651879 \\ \hline
\end{tabular}
\end{table*}

\begin{figure}[H]
 \centering
 \includegraphics[width=.5\textwidth]{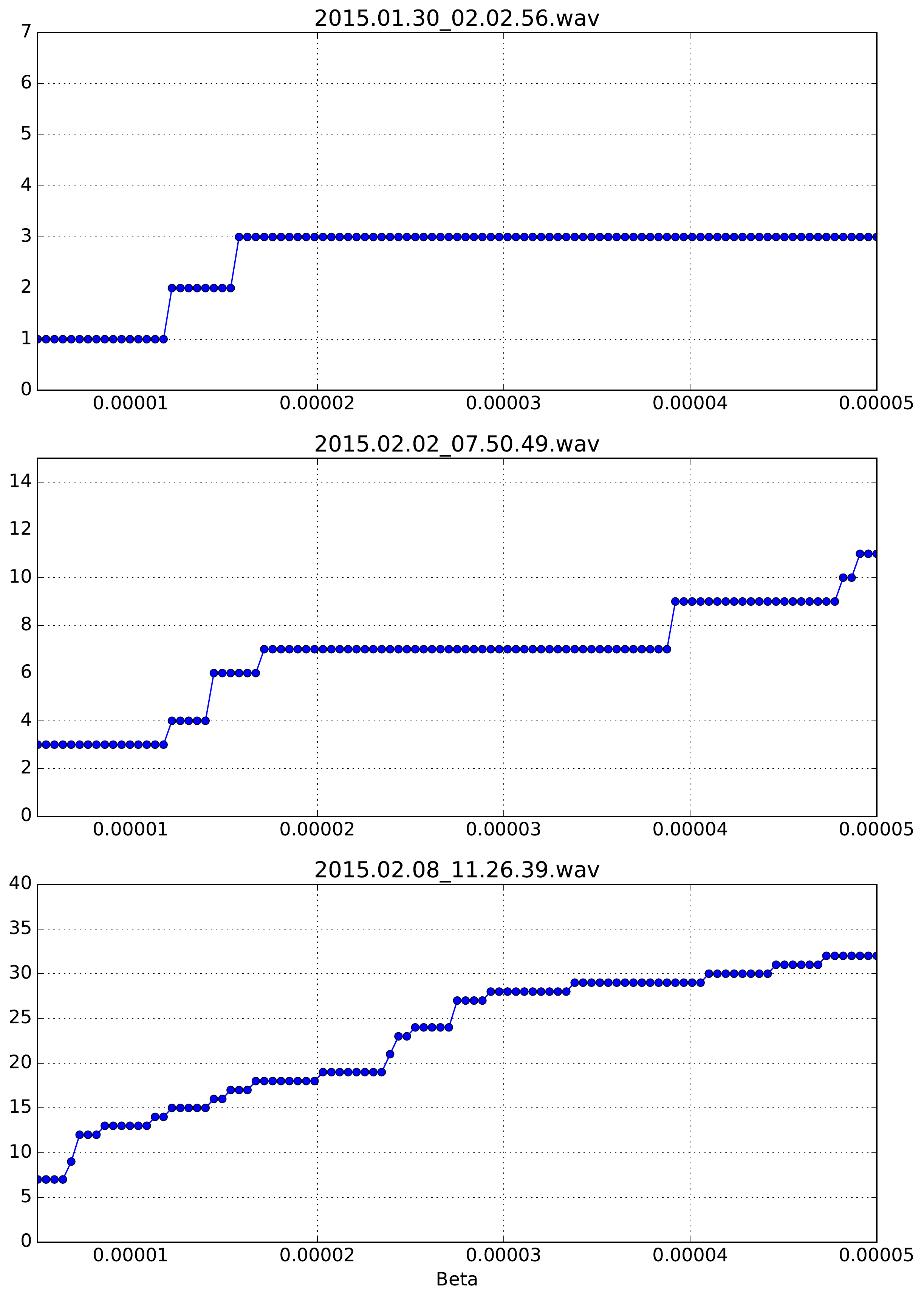}
 \caption{Segmentation of three signals}
 \label{fig:seg}
\end{figure}

The first thing we note is that indeed the execution times dropped sensibly: all segmentation tasks shows in table \ref{tbl:table7} took less than 2 seconds. This is a major accomplishment for our algorithm, one that allows us to experiment with it and to apply it to the segmentation of very long signals, in the order of months or even years.

However, we also noticed a high sensibility for $\beta$ on the number of segments found, specially in the third sample that contains many events. The number of segments grows with $\beta$, and shows little stabilization. Also, if we pick a value for $\beta$ with high accuracy on the first two samples, we will ignore many meaningful segments that could be otherwise found in the third sample. In other words, there might not exist a single optimal value for $\beta$ that will work for all kinds of signals. 

One posible way to deal with that is to choose $\beta$ by balancing the analysis's goals: one might consider what is preferable, to have many segments of a uniform signal, or to have longer segments with possibly other change points going undetected inside them. It is really a matter of the signal's diversity and of what the subsequent step of the analysis is going to be. Consider, for instance, that after segmentating the signal one applies a clustering procedure to group together signals with similar features; in this case, it might be best to choose a higher value of $\beta$, allowing some oversegmentation and expecting that these uniform segments would be again classified together after clusterization. 

Another strategy might be to start the segmentation procedure with a low $\beta$, increase it by a small factor and run the segmentation again. One can then accept the segmentation when there is some degree of stabilization (for instance if the number of segments is the same after $10$ steps with increasing $\beta$ values). Of course, this kind of procedure would add two new parameteres to the main algorithm: the increase factor for $\beta$, and the stabilization parameter. 

\begin{figure*}
 \centering
 \includegraphics[width=.9\textwidth]{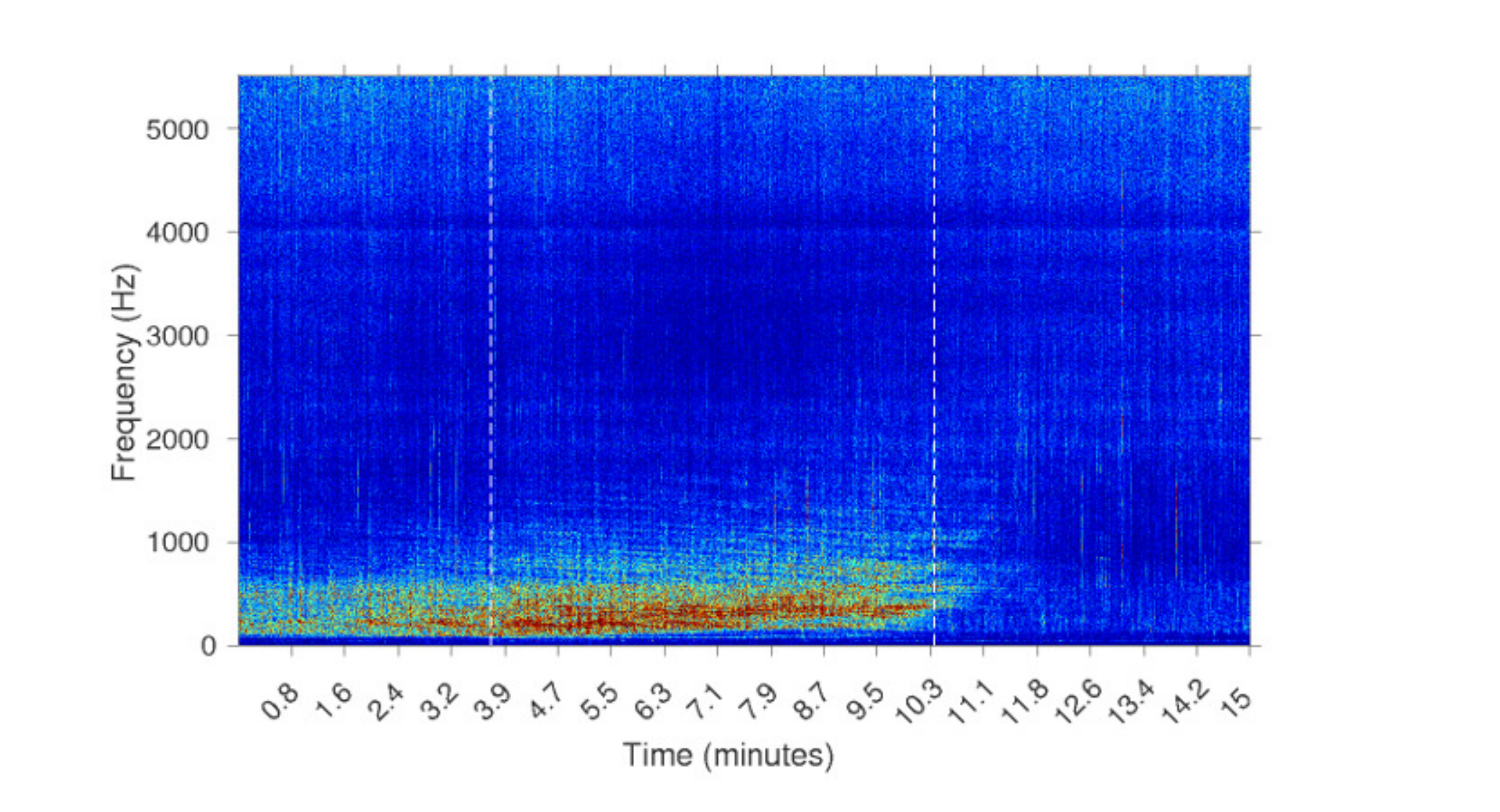}
 \caption{Segmentation of the long event}
 \label{fig:res0202}
\end{figure*}

\begin{figure*}
 \centering
 \includegraphics[width=.9\textwidth]{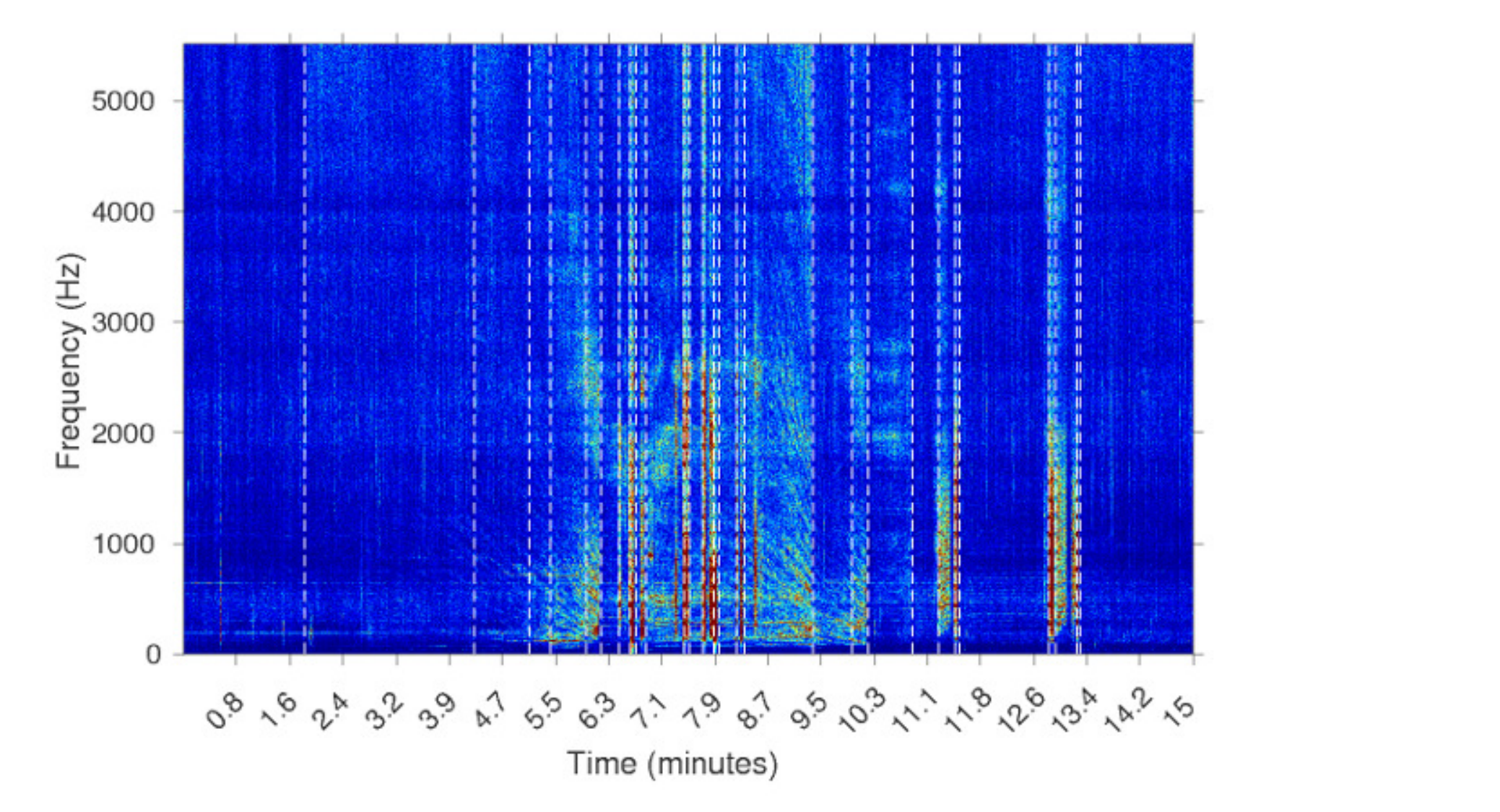}
 \caption{Segmentation of many short duration events}
 \label{fig:res0802}
\end{figure*}

In the samples we analyzed in this section, we can see what would be the outcome of this method: if we started at $\beta = 5e^{-6}$, as we did, increased it by an additive factor of $4.5e^{-7}$, and waited for $10$ consecutive runs of the algorithm with the same number of segments, we would end with $1$ segments for the first sample, $3$ for the second and $29$ for the third. This numbers seem quite reasonable, as we can see from figures \ref{fig:res0202} and \ref{fig:res0802}.

A third possibility would be to propose an adaptive scheme for $\beta$, and increase it as the segment sizes become smaller. Again, this would include new parameters in the algorithm, and also would demand an analysis of the precise form that $\beta$ should take as function of segment sizes.

\section{Conclusions}\label{sec:conc}

The sequential segmentation algorithm is a Bayesian unsupervised methodology aimed at segmentating audio signals. It is based on the basic assumption that the occurrence of events in an audio signal induces a change in the signal's total power. 

In a previous work we presented the algorithm and compared it to a peak detection strategy, with promising results. However, the computing time involved in the sequential segmentation algorithm was almost prohibitive; also, because of this high computing cost, it was not possible to provide a detailed study of the algorithm's sensibility to parameters.

In this paper we presented a Python module that implements the sequential segmentation algorithm using Cython. With this new implementation, we were able to reduce the computing time from $\approx 120s$ to $\approx 1s$ in the same conditions as before. This represents a performance gain in the order of $100$ times. 

We have also provided details on the MCMC step involved in the calculation of evidence values for the hypothesis of equality of variances, which is part of the algorithm's stopping criteria, and finally we analyzed the algorithm's sensibility to its parameters using both simulated and real signals.

What we found is that the algorithm is robust to the acceptance threshold, $\alpha$ in most situations; we recommend setting $\alpha = 0.1$ as a default value. Once fixed the value of $\alpha$, there remains only the need for calibrating for $\beta$. 

The calibration procedure for this parameter, however, depends on the characteristics of the data and the goals of the analysis.

The code and data files used in this paper are available at \url{https://github.com/paulohubert/bayeseg}.


\end{document}